\documentclass[%
 reprint,
 amsmath,amssymb,
 pra,superscriptaddress
 ]{revtex4-1}

\usepackage{mypackages}

\begin{document}

\title{A fault-tolerant variational quantum algorithm with limited \textit{T}-depth}

\author{Hasan Sayginel}
\email{hasan.sayginel.17@ucl.ac.uk}
\affiliation{Department of Physics and Astronomy, University College London, London, WC1E 6BT, United Kingdom}
\affiliation{National Physical Laboratory, Teddington, TW11 0LW, United Kingdom}
\author{Francois Jamet}
\affiliation{National Physical Laboratory, Teddington, TW11 0LW, United Kingdom}
\author{Abhishek Agarwal}
\affiliation{National Physical Laboratory, Teddington, TW11 0LW, United Kingdom}
\author{Dan E. Browne}
\affiliation{Department of Physics and Astronomy, University College London, London, WC1E 6BT, United Kingdom}
\author{Ivan Rungger}
\email{ivan.rungger@npl.co.uk}
\affiliation{National Physical Laboratory, Teddington, TW11 0LW, United Kingdom}

\date{\today}

\begin{abstract}
We propose a variational quantum eigensolver (VQE) algorithm that uses a fault-tolerant gate-set, and is hence suitable for implementation on a future error-corrected quantum computer. VQE quantum circuits are typically designed for near-term, noisy quantum devices and have continuously parameterized rotation gates as the central building block. On the other hand, a fault-tolerant quantum computer can only implement a discrete set of logical gates, such as the so-called Clifford+\textit{T} gates.
We show that the energy minimization of VQE can be performed with such a fault-tolerant discrete gate-set, where we use the Ross-Selinger algorithm to transpile the continuous rotation gates to the error-correctable Clifford+\textit{T} gate-set. We find that there is no loss of convergence when compared to the one of parameterized circuits if an adaptive accuracy of the transpilation is used in the VQE optimization. State preparation with VQE requires only a moderate number of \textit{T}-gates, depending on the system size and transpilation accuracy. We demonstrate these properties on emulators for two prototypical spin models with up to $16$ qubits. This is a promising result for the integration of VQE and more generally variational algorithms in the emerging fault-tolerant setting, where they can form building blocks of the general quantum algorithms that will become accessible in a fault-tolerant quantum computer.
\end{abstract}

\maketitle


\section{Introduction}
\label{sec:intro}
Quantum computing has seen significant progress in the last decade, leading to the development of quantum processors with an increasing number of qubits. These noisy intermediate-scale quantum (NISQ) processors are currently being investigated for a possible advantage over classical computers \cite{google_supremacy,advantage_1,advantage_2,advantage_3}. However, anticipated general quantum algorithms with a proven exponential speed-up over their classical counterparts lie outside of the reach of NISQ processors, because of the limited physical qubit numbers as well as the accumulation of noise for deep circuits \cite{howtofactorbit, Kivlichan2020improvedfault, Campbell_2022}. Such algorithms include Shor's algorithm \cite{shor} and quantum phase estimation (QPE) \cite{phase_estimation}. Therefore, for the full potential of quantum computing to be achieved, fault-tolerant quantum computers will be needed.

An essential part of developing a fault-tolerant quantum computer (FTQC) is the incorporation of quantum error correction (QEC) into the processors, which will not only increase the coherence times of the resulting logical qubits but also actively correct for errors in faulty operations as the information is being processed. To this end, there have been recent prototype demonstrations of QEC on superconducting \cite{qec_superconductor} and ion-trap devices \cite{qec_ion}. Furthermore, recently error suppression with increasing code-size was demonstrated for the surface code \cite{google_qec_errorsupress} despite an increased number of physical qubits and gate operations. While the gate error levels in current hardware are not yet sufficiently low for large-scale error correction, with rapid progress in the field,  error-corrected quantum computers with a small number of logical qubits may emerge in the next few years. 

 The main difference between the quantum circuits achievable on near-term, so called noisy-intermediate scale quantum (NISQ) computers, and future error-corrected quantum computers are the range of quantum gates available at the logical level. Typically, NISQ devices allow continuously parameterized single-qubit rotations, such as an arbitrary $z$-rotation, $R_z(\theta)$, where $\theta\in[0,2\pi]$. On the other hand, FTQCs only allow a fixed set of discrete rotation angles, such as the $\pi/4$ rotation about the $z$-axis, also known as the \textit{T}-gate. For this reason, any algorithm must first be compiled into a code-specific fault-tolerant gate-set before it can be run on the logical qubits. Fault-tolerant gate synthesis has been investigated in depth, and a number of algorithms have been proposed \cite{solovay_kitaev_pedagogical, ross-selinger,cliffordT_1,new_T_count_paper,cliffordT2,cliffordT_3}. 

In this article we study the variational quantum eigensolver (VQE) algorithm \cite{vqe_original} in the context of fault-tolerant quantum computation. VQE uses the variational principle to compute the ground state energy of a Hamiltonian, a task that commonly arises in condensed matter physics and quantum chemistry \cite{TILLY2022}. Due to its partial resilience to noise for moderate numbers of qubits and circuit depths, VQE has allowed to perform a number of proof of concept demonstrations on quantum hardware. However, the scaling to larger systems is hindered by the noise in the hardware, since the noise mitigation methods used successfully in small experiments become very expensive to scale up to larger circuits \cite{TILLY2022,noise_mitigation}. Therefore, fault-tolerant computations using QEC could be a way to realize larger-scale VQE. We note that FTQC would not solve all the intrinsic scaling challenges of VQE, such as the difficulty in the classical optimization of the circuit parameters for large systems \cite{TILLY2022,barren_plateaus,noise_induced_barren_plateus}. However, in the fault-tolerant setting, we expect VQE to be used alongside algorithms with known scalability such as QPE \cite{phase_estimation}. For example, VQE may be used as a way to prepare an initial state with enough overlap with the ground state for a QPE algorithm to perform more efficiently.

The hurdle which must be overcome is that current VQE quantum circuits, designed for near-term hardware, include continuously parameterized rotation gates. The rotation angles are optimized on a classical computer, and the energies used in the optimization process for a given set of angles are obtained from the quantum computer. It had not yet been investigated how well this classical optimization process can  be performed given the limitations of a fault-tolerant discrete gate-set, where a continuous range of parameters can only be achieved in the limit of large numbers of $T$-gates, and whether the necessary approximations in the gates for circuits with finite $T$-gate depth would affect the converge of the algorithm. In this article, we address this question and show that by integrating the Ross-Selinger recompilation of a continuous rotation to an FT gate-set into our algorithm, we can obtain systematic convergence of the VQE algorithm. We find that there is no slowdown in convergence efficiency when compared to the conventional VQE circuits with parameterized hardware gates if an adaptive Ross-Selinger recompilation accuracy is used.

The structure of the paper is as follows. Sec. \ref{method} discusses the methods, introducing the fault-tolerant Clifford+\textit{T} gate-set, the Ross-Selinger (RS) algorithm used for fault-tolerant gate synthesis, and a fault-tolerant implementation of the VQE algorithm.  Sec. \ref{results} introduces the two spin models that we use to test the method, and then the results for both fixed angles of a quantum circuit and for a full VQE loop performed using our implementation with an FT gate-set. In Sec. \ref{conclusions} we discuss the conclusions.  

\section{methods}\label{method}

\subsection{Fault-tolerant gate-set}\label{gateset}
The universal fault-tolerant gate-set used in this study is the Clifford+\textit{T} gate-set. The Clifford set consisting of hadamard, phase-gate, and controlled-NOT gates, $\{H,S,\text{CNOT}\}$, which can generate any $N$-qubit Clifford operation. With the addition of the $T$-gate, any unitary lying in $\text{SU}(2^N)$ can be approximated up to any target accuracy by a sequence of Clifford+\textit{T} gates \cite{NeebRainesSloane}.  
The \textit{T}-gate is expressed in matrix form as 
\begin{equation}
    T = \mqty[1 & 0 \\ 0 & e^{i\pi/4}].
\end{equation}
 
 There are diverse ways of achieving fault-tolerant logic gates depending on the underlying code. However, in many codes, including the surface code \cite{cite_surface_code} (currently the most promising quantum error correcting code due to its high threshold and simple two-dimensional modular implementation), Clifford group gates can be performed in a direct and fast method, while $T$-gates are more expensive. 
 
 In order to implement the \textit{T}-gates in these codes, a non-unitary technique such as magic-state distillation and injection is required. This procedure has a spatial and temporal overhead  due to the additional ancillary qubits and physical gates needed \cite{game_of_surface_codes}. Therefore, an important metric in the analysis of the performance of a fault-tolerant quantum algorithm implemented with Clifford+\textit{T} gates is the number of \textit{T}-gates. 
 

In our work, we are, therefore, interested in the two following metrics. 
\begin{itemize}
    \item \textbf{\textit{T}-count}: \textit{T}-count is defined as the total number of \textit{T} gates in the fault-tolerant quantum circuit.
    \item \textbf{\textit{T}-depth}: \textit{T}-depth is defined as the total number of layers of \textit{T} gates, where within each layer parallel execution of the \textit{T} gates on different logical qubits is possible.
\end{itemize}

\subsection{Ross-Selinger algorithm}
The Ross-Selinger (RS) algorithm computes approximations of arbitrary single qubit z-rotations, $R_z(\theta)$ over the Clifford+\textit{T} gates. It achieves this by approximating $R_z(\theta)$ with another unitary $U$, where $U$ has an exact decomposition over Clifford+\textit{T} gates up to the single-qubit global phase, $\omega = e^{i\frac{\pi}{4}}$ \cite{ross-selinger}.
\small
\begin{equation}
    R_z(\theta) = \mqty[e^{-i\frac{\theta}{2}} & 0 \\ 0 & e^{i\frac{\theta}{2}}] \xrightarrow{\text{RS}} U = \prod_m U_m\in\{\omega,H,S,T\}
    \label{eq:rs1}
\end{equation}
\normalsize
The algorithm then outputs the Clifford+\textit{T} decomposition of the unitary $U$. The error in this approximation is given as 
\begin{equation}
    ||R_z(\theta)-U||\leq\epsilon=10^{-d}
    \label{eq:norm}
\end{equation}
where $||.||$ is the operator norm bounded by $\epsilon$, and $d$ is the digit accuracy. The accuracy of the decomposition can be systematically improved with increasing \textit{T}-depth. Furthermore, the Ross-Selinger algorithm is efficient in the number of \textit{T}-gates, which in the typical case scales with respect to $\epsilon$ as $4\log_2(\frac{1}{\epsilon}) + O\big(\log(\log(\frac{1}{\epsilon}))\big)$. This is in contrast to the Solovay-Kitaev algorithm \cite{solovay_kitaev_pedagogical}, which achieves \textit{T}-counts of $O\big(\log^c(\frac{1}{\epsilon})\big)$, where $c$ is a constant, making Ross-Selinger more favorable for $c>3$. 

Rotations about other axes, such as $R_x$ and $R_y$, can be achieved with the corresponding $R_z$ rotation and additional Clifford gates as
\begin{equation}
    R_x(\theta) = HR_z(\theta)H, \;\;\; R_y = SHR_z(\theta)HS^\dagger. \label{eq:rs2}
\end{equation}
Therefore, the Ross-Selinger transpilation can be used to find the Clifford+\textit{T} approximation of any $\text{SU(2)}$ unitary. 

\begin{figure}[H]
    \centering
    \includegraphics[scale = 0.42]{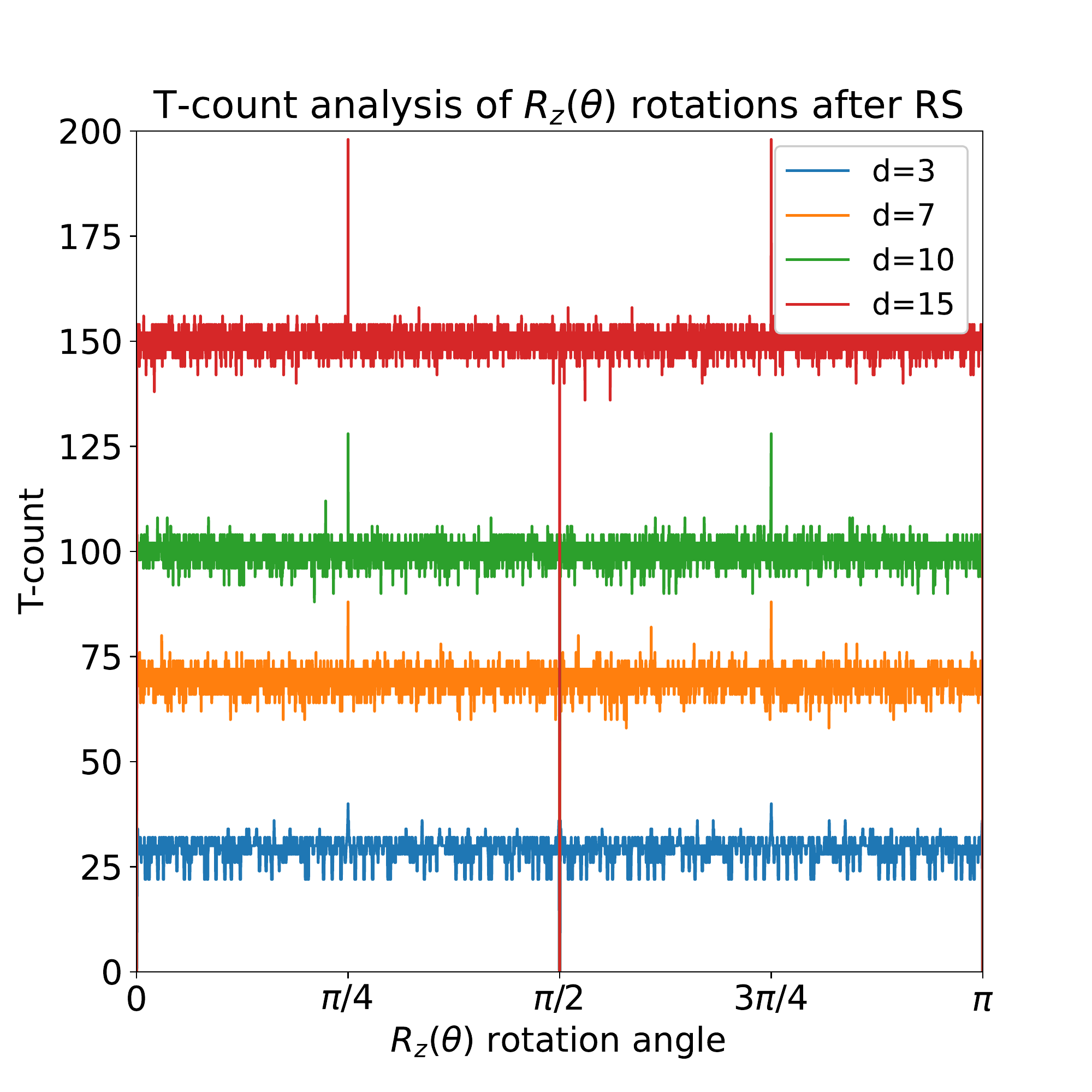}
    \caption{\textit{T}-counts of $R_z(\theta)$ rotations for increasing RS digit accuracy, $d$ (see Eq. \ref{eq:norm}).}
    \label{fig:rz_rots_t_depth}
\end{figure}

The software implementation of the RS algorithm is available in an open-source software package \cite{newsynth}. 
Fig. \ref{fig:rz_rots_t_depth} shows practical numbers for the number of \textit{T} gates in the RS decomposition of $R_z(\theta)$ rotations in the range $\theta\in[0,\pi]$ at various $d$, obtained using this software package. It can be observed that for a given $d$, the number of \textit{T} gates required to realize different rotations is similar for all rotations. Note that multiples of $\pi/2$ can be carried out by \textit{S} gates and hence  have a \textit{T}-count of $0$.



\subsection{Fault-tolerant implementation of VQE}
\label{vqe}

The VQE is a hybrid quantum-classical algorithm, where the ground state wave function of a Hamiltonian $\mathcal{H}$ is expressed on the quantum computer by executing a parametric quantum circuit, denoted as ansatz $V(\boldsymbol{\theta})\ket{0}$.  $V(\boldsymbol{\theta})$ is a sequence of parametric gates, and $\boldsymbol{\theta}$ is a vector of parameters that is optimized to find the ground state. The energy of the parametric wave function,
\begin{equation}
    \mathcal{E}(\boldsymbol{\theta}) = \mel{0}{V(\boldsymbol{\theta})^\dagger \mathcal{H} V(\boldsymbol{\theta})}{0},
    \label{eq:loss}
\end{equation}
can be obtained on a quantum processor as expectation value.

 If the quantum circuit can span the entire Hilbert space, the minimum value of $\mathcal{E}(\boldsymbol{\theta})$ will correspond to the ground state of $\mathcal{H}$. However, practical implementations typically use a circuit with limited parameters, leading to an approximation of the ground state. The minimum value of $\mathcal{E}(\boldsymbol{\theta})$ is found via classical optimization.

\subsubsection*{Classical optimization}
The task of finding the minimum value of $\mathcal{E}(\boldsymbol{\theta})$ is challenging, because it is usually a non-convex function. In this paper we focus on gradient-based minimization methods, specifically the Broyden-Fletcher-Goldfarb-Shanno (BFGS) algorithm \cite{broyden,fletcher,goldfarb,shanno}. In BFGS, the procedure assumes a continuous optimization landscape as a function of the parameters. On the contrary, in fault-tolerant quantum computation, one will necessarily be  restricted to a discrete gate-set, typically Clifford+T. Here, we therefore substitute the parametric gate-set $R_x(\theta),R_y(\theta),R_z(\theta)$ by their RS decompositions (Eqs. \ref{eq:rs1} and \ref{eq:rs2}).


Two strategies are usually used to obtain the gradient in VQE: the finite-difference rule and the parameter-shift rule \cite{quantum_circuit_learning,general_param_rule,gavin_crooks_param}. In this paper, we use both strategies and evaluate their relative performance.



\underline{\textit{Finite-difference rule}}: this method requires two evaluations of a given circuit with an infinitesimally shifted parameter. The required derivatives can be approximately calculated as
\small
\begin{align}
    \partial_{\theta_\mu}\mathcal{E}(\theta_\mu)
    \approx
     \frac{\mathcal{E}(\theta_\mu + \frac{1}{2}\Delta\theta_\mu) - \mathcal{E}(\theta_\mu - \frac{1}{2}\Delta\theta_\mu)}{\Delta\theta_\mu}.
\end{align}
\normalsize
 For ease of notation in the equation we only specify the shifted parameter $\theta_\mu$  as argument in the function; all other parameters in the vector $\boldsymbol{\theta}$ are kept constant. For the fault-tolerant VQE algorithm, the finite-shift rule requires two runs of the Ross-Selinger algorithm for calculating the Clifford+\textit{T} decompositions of the shifted angles, followed by two expectation value evaluations.

\underline{\textit{Parameter-shift rule}}: the parameter-shift rule is a method of computing the gradient of a parameterized gate by running the same quantum circuit twice with a finite shift in the gate parameter \cite{quantum_circuit_learning,general_param_rule,gavin_crooks_param}. It states that if the generator $G$ of the gate $\mathcal{G}(\theta_\mu)$ (with a single parameter $\theta_\mu$) has two unique eigenvalues denoted as $e_0$ and $e_1$, then the derivative of the circuit expectation with respect to the gate parameter $\theta_\mu$ is given by the difference in expectation value of two circuits. The two circuits are run with shifted parameters scaled by a parameter $r$ given as $r = \frac{e_1 - e_0}{2}$. The gradient is obtained as
\begin{equation}\label{eq:param_shift}
    \partial_{\theta_\mu} \mathcal{E}(\theta_\mu) = r\Big[\mathcal{E}\Big(\theta_\mu + \frac{\pi}{4r}\Big) - \mathcal{E}\Big(\theta_\mu - \frac{\pi}{4r}\Big)\Big]
\end{equation}
\noindent where $r$ is the scaling parameter and $\frac{\pi}{4r}$ is a finite shift.

The main gate of interest in this work is $R_z(\theta) = e^{-i\frac{\theta}{2}Z}$. The \textit{Z}-gate has eigenvalues $\pm1$, hence the value of $r$ for $R_z(\theta)$ is $r=\frac{1}{2}$. The parameter-shift rule for $R_z(\theta)$ is, therefore, expressed as
\begin{equation}\label{eq:param_shift_rz}
    \partial_{\theta_\mu} \mathcal{E}(\theta_\mu) = \frac{1}{2}\Big[\mathcal{E}\Big(\theta_\mu + \frac{\pi}{2}\Big) - \mathcal{E}\Big(\theta_\mu - \frac{\pi}{2}\Big)\Big].
\end{equation}
We note that the required parameter-shift can be exactly implemented with $S$ gates together with the appropriate multiple of the phase $\omega$:
$R_z(\pi/2) = \omega^7S$ and $R_z(-\pi/2) = \omega SSS$.

We remark that the two-qubit gate $ZZ(\theta) = e^{i\frac{\theta}{2}Z\otimes Z}$, which is required in the Hamiltonian Variational Ansatz used in the results section of this article, can be decomposed into CNOTs and a single-qubit rotation. The gradient of the $ZZ(\theta)$ can therefore be calculated with Eq. \ref{eq:param_shift_rz}. 
\begin{figure}[H]
    \centering
    \includegraphics[scale=0.060]{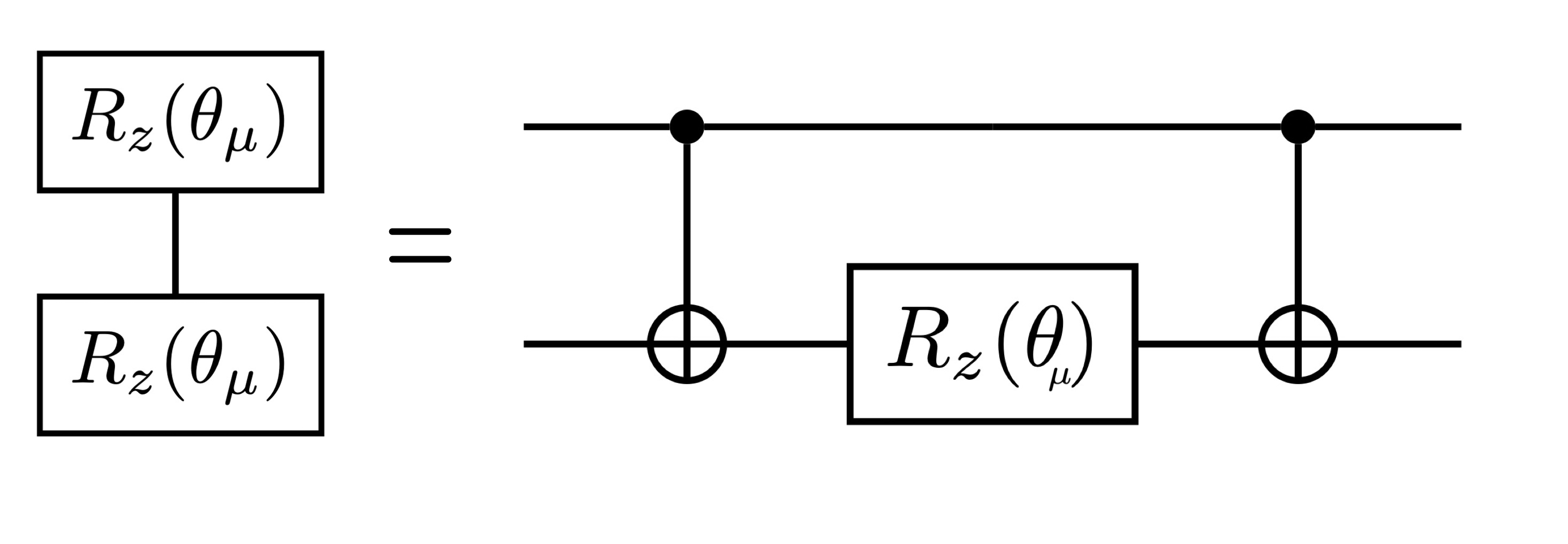}
    \caption{$ZZ(\theta)$ gate decomposed into CNOTs and a single-qubit $R_z(\theta)$ gate.}
    \label{fig:zz_theta}
\end{figure}
\noindent Therefore, the parameter-shift rule for $ZZ(\theta)$-like gates does not have an additional fault-tolerant overhead, i.e., no extra \textit{T}-gates are needed for the required shifts because \textit{S} gates can realize the desired shifts.

\section{Results}\label{results}
To demonstrate the method, we apply the FT-VQE to two spin models, namely the transverse-field Ising model (TFIM) and the XXZ model, using a Hamiltonian Variational Ansatz (HVA) \cite{hva_original,hva_new}. 
Within the HVA each ansatz is model-specific. For a general Hamiltonian, $\mathcal{H}$, which is a linear sum of not-necessarily commuting terms $\mathcal{H} = \sum_i \mathcal{H}_i$, an HVA ansatz is expressed as
\begin{equation}
    \ket{\psi_L} = \prod_{l=1}^{L}\Big(\prod_i \exp(-i\theta_{i,l}\mathcal{H}_i)\Big)\ket{\psi_0} 
\end{equation}
\noindent where $L$ is the total number of layers (layer depth). $\ket{\psi_0}$ is the ground state of one of the individual terms $\mathcal{H}_i$ of the Hamiltonian \cite{hva_original,hva_new}. HVA is motivated by the Quantum Approximate Optimization Algorithm \cite{qaoa}, and the choice of $\ket{\psi_0}$ is similar, which is a state that is easy to prepare in hardware.
For our investigation, HVA is chosen because of its linear scaling in the number of parameters with increasing number of layers.

\begin{figure}
    \centering
    \includegraphics[scale=0.04]{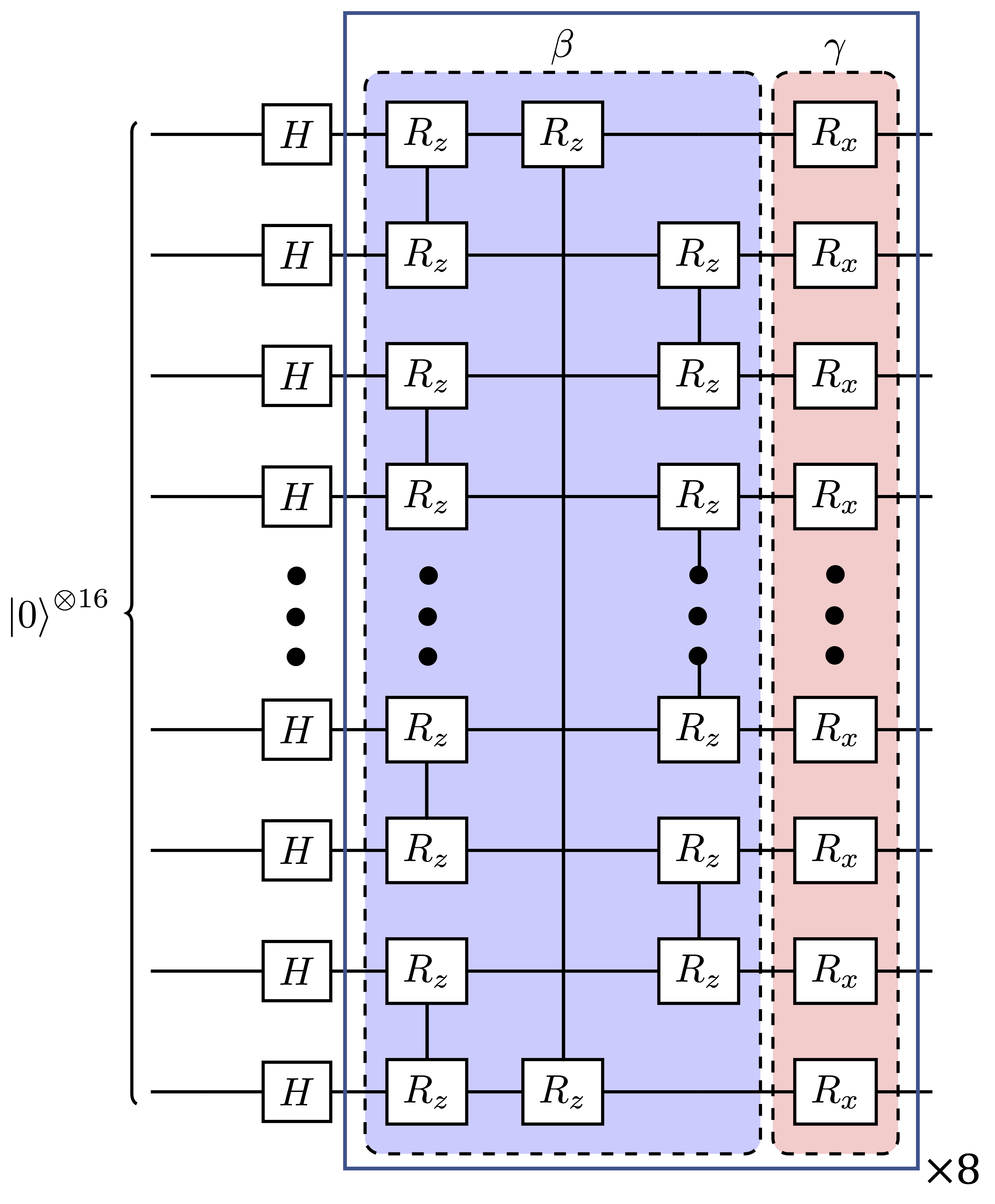}
    \caption{TFIM ansatz with $16$ qubits with 2 parameters per layer, where $\beta,\gamma$ are the variational parameters of all the gates in each layer. Layers are repeated $8$ times leading to $16$ free parameters in total.}
    \label{fig:tfim_ansatz}
\end{figure}

\underline{\textit{Transverse-field Ising model (TFIM)}}. The TFIM Hamiltonian is a prototype model of quantum magnetism, and is given by
\begin{equation}
    \mathcal{H}_{\text{TFIM}} = - \sum_{i=1}^{N}[ \hat{Z}_{i} \hat{Z}_{i+1} + g \hat{X}_i],
\end{equation}
where we consider periodic boundary conditions ($\hat{Z}_{N+1} = \hat{Z}_{1}$). The value of $g$ affects the magnetic phase, where $g<1$ gives a ferromagnetic phase as ground state, $g>1$ corresponds to a paramagnetic phase, and $g=1$ is a critical point. The system is gapless at $g=1$ in the thermodynamic limit where $N$ goes to infinity. For our FT-VQE simulations, we consider the critical $g=1$ point for a system of $N=16$ qubits. The initial state is taken as $\ket{\psi_0}=\ket{+}^{\otimes 16}$. For the HVA we use a layer depth of $L=8$, leading to $16$ parameters. The corresponding quantum circuit ansatz is shown in Fig.\ \ref{fig:tfim_ansatz}. This system size and ansatz choice are based on those of Ref. \cite{hva_new}, where the entanglement and optimization properties of this system have been first investigated using the HVA. In Ref. \cite{hva_new} it is shown that barren plateaus are expected at $L=8$ with a random parameter initialization. We, therefore, choose this setup as one of our test systems to evaluate the performance of FT-VQE in presence of barren plateaus.

\begin{figure}
    \centering
    \includegraphics[scale = 0.40]{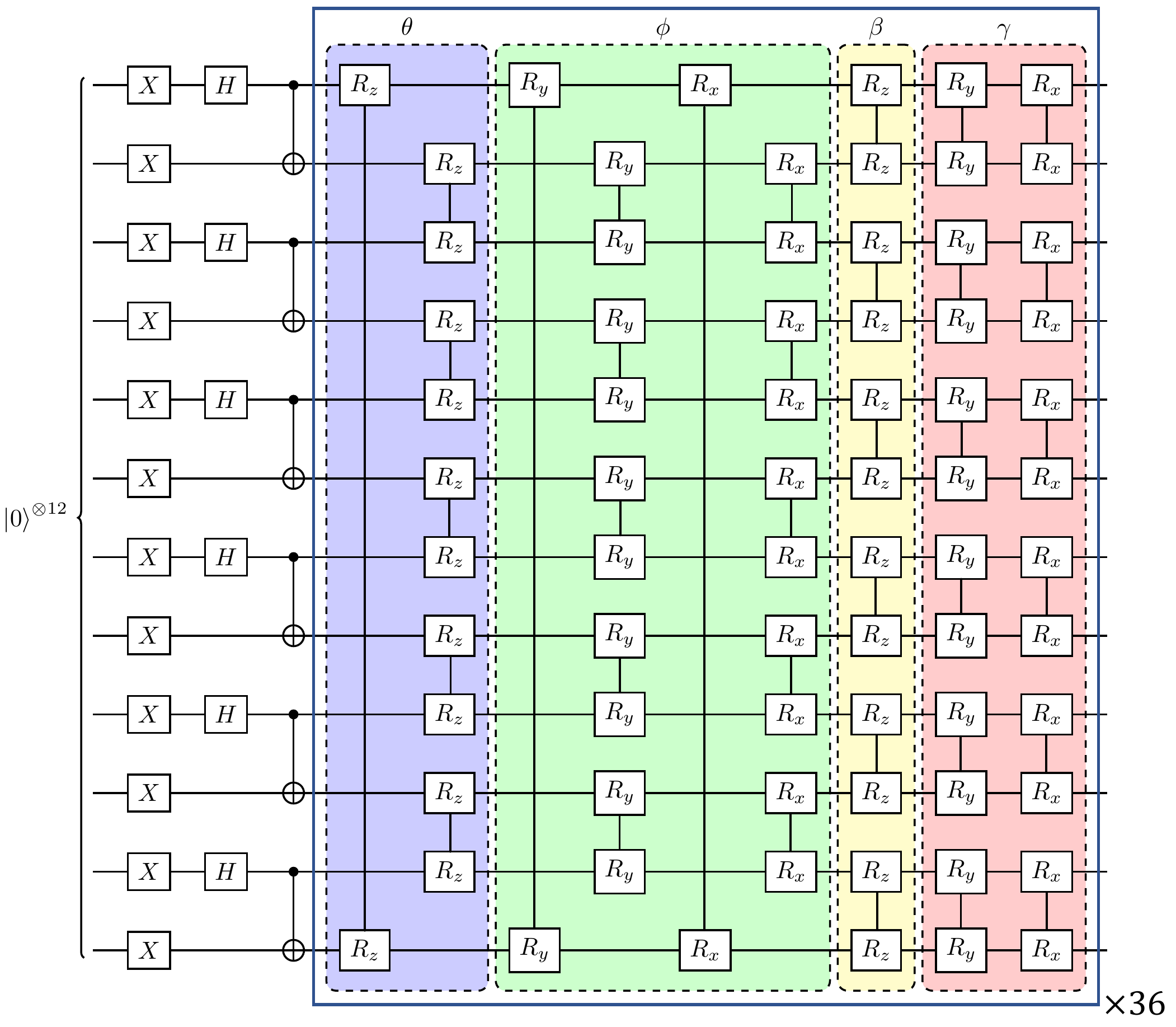}
    \caption{XXZ model ansatz with $12$ qubits and 4 parameters per layer, where $\theta,\phi,\beta,\gamma$ are the variational parameters of all the gates in each layer. Layers are repeated $36$ times leading to $144$ free parameters in total. It was numerically observed that at least $36$ layers were needed for sufficient convergence with random initial parameters.}
    \label{fig:xxz_ansatz}
\end{figure}

\underline{\textit{XXZ model}}. The XXZ model is another prototypical model of quantum magnetism. In the one-dimensional case its Hamiltonian is given by
\begin{align}
        \mathcal{H}_{\text{XXZ}} = \sum_{i=1}^{N} \hat{X}_{i} \hat{X}_{i+1} + \hat{Y}_{i} \hat{Y}_{i+1} + \Delta \hat{Z}_{i} \hat{Z}_{i+1} ,
\end{align}
\noindent where again periodic boundary conditions are applied. $\Delta$ represents the spin anisotropy of the model. For our FT-VQE simulations, we take $\Delta=1$, at which there is a phase transition to the N\'eel ordered state. The system size we consider is $N=12$ qubits, and for the HVA ansatz we use a layer depth $L=36$, leading to $144$ free parameters. The initial state is taken as $\ket{\psi_0}= \bigotimes_{n=1}^{N=6}\ket{\Psi^-}$, where $\ket{\Psi^-}=\frac{1}{\sqrt{2}}(\ket{01}-\ket{10})$ is a Bell state as in \cite{hva_new}. The ansatz is shown in Fig. \ref{fig:xxz_ansatz}. In Ref. \cite{hva_new} it is shown that such ansatz achieves good VQE convergence at $L=36$, which is attributed to the system being in the over-parameterization limit for the XXZ model, and hence avoids a barren plateau. Therefore, this second ansatz illustrated in Fig. \ref{fig:xxz_ansatz} allows us to evaluate the performance of FT-VQE for a system that has no barren plateaus, but a large number of variational parameters and layers.

Finally, all the calculations performed in this paper have been done on the quantum emulator Qulacs \cite{qulacs} as state vector emulations.

\subsection{State preparation accuracy analysis with limited \textit{T}-depth}

\begin{figure}
    \centering
    \includegraphics[scale = 0.37]{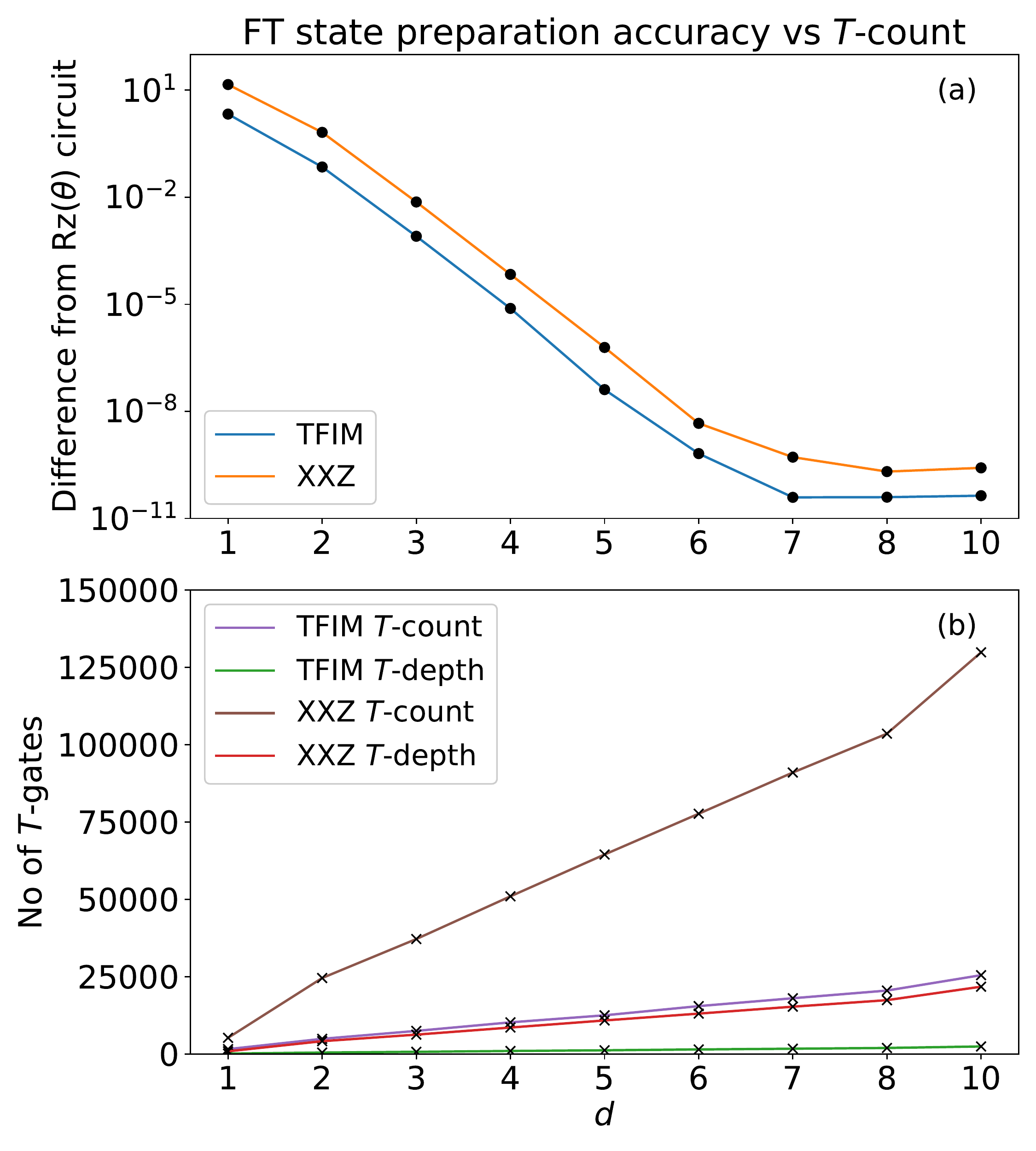}
    \caption{Difference of the ground state energy between the solution obtained with the circuit using parameterized rotations (Rz($\theta$)-circuit) and its fault-tolerant form with limited \textit{T}-depth following Clifford+\textit{T} compilation (FT($\theta$)-circuit). The horizontal axis of the plots illustrates the Ross-Selinger digit accuracy, $d$ in Eq. \ref{eq:norm}. (a) Difference in energy of the Rz($\theta$)-circuit and the FT($\theta$)-circuit for the same rotation parameters, as function of digit precision $d$ (b) \textit{T}-count and \textit{T}-depth, showing that they increase approximately linearly with $d$.}
  \label{fig:t_depths}
\end{figure}

First, we perform the VQE minimization using the parameterized rotation gates. We denote this circuit as the Rz($\theta$)-circuit due to the continuous parameterization of the rotation angles, and due to the decomposition of each parameterized gate into single-qubit $R_z(\theta)$ rotations. We then fix the rotation gates at the optimized values, perform the RS decomposition at different $d$ for each rotation gate, and then replace all parameterized $R_z(\theta)$ rotations with their RS decompositions. We denote this circuit as the FT($\theta$)-circuit. We evaluate the energy difference between the Rz($\theta$)-circuit and FT($\theta$)-circuits for different digital accuracy $d$ (Eq. \ref{eq:norm}). Fig. \ref{fig:t_depths} shows this comparison for both the TFIM and XXZ models, together with the resulting $T$-count and $T$-depth. The energy difference decreases exponentially with $d$, until $d$ is larger than about 6, above which the energy difference becomes approximately constant. We verified that this tail-off is due to the finite numerical accuracy of the emulator software. In Appendix \ref{mathematica} we show that when performing the numerical evaluations at higher precision the energy difference decays exponentially up to digit accuracy of $d=16$. Our results therefore show that one can systematically increase the accuracy of the energy of the FT($\theta$)-circuit by increasing $d$ up to the numerical accuracy. Importantly, the circuit depth only increases linearly with $d$, as guaranteed by the RS decomposition. 

For $d{=}4$ the error in the energy expectation value is less than $10^{-4}$ for both the TFIM and XXZ models. For the TFIM model the corresponding \textit{T}-count is $10232$, and the \textit{T}-depth is $974$; for the XXZ model the \textit{T}-count is $50976$ and the \textit{T}-depth is $8554$. The energy difference becomes approximately constant for $d$ larger than 6, where it reaches values of the order of 10$^{-10}$ to $10^{-9}$. The \textit{T}-count for $d=6$ is $15480$, with a \textit{T}-depth of $1474$, for the TFIM; for the XXZ model the \textit{T}-count is $77680$, with a \textit{T}-depth of $13056$. TFIM with $N=16$ qubits has a total number of $256$ single-qubit $R_z(\theta)$ gates, whereas the XXZ model has a total of $1296$ single-qubit $R_z(\theta)$ gates. The total \textit{T}-count for both models at arbitrary $d$ is approximately equal to the number of $R_z(\theta)$ gates multiplied by the average number of \textit{T}-gates per RS decomposition of a single $R_z(\theta)$ gate, which can be extracted from Fig. \ref{fig:rz_rots_t_depth}. Note that in practice also the $R_z(\theta)$ gates have finite accuracy when implemented in hardware, so that the hardware result will also deviate from the ideal exact circuit results.

The above \textit{T}-count estimates are modest when compared with recent resource count estimates for other fault-tolerant algorithms. For example, for the problem of integer factoring, it is expected that at least $10^9$ \textit{T}-gates and $10^5$ logical qubits will be needed \cite{howtofactorbit}. In Refs. \cite{t_count1, t_count2} the required $T$-counts of various quantum simulation problems are investigated, and estimates range between $10^{7}-10^{12}$ \textit{T}-gates depending on the complexity of the system. Particularly, in Ref. \cite{t_count2}, authors estimate the \textit{T}-counts of various spin systems. The Hamiltonian studied in \cite{t_count2} is similar to our XXZ model with $\Delta=1$ (a Heisenberg chain), and an additional transverse-magnetic field. Their lower estimates are of the order of $10^7$ \textit{T}-gates for a spin-system with $14$ qubits for an error level of $10^{-3}$. For the XXZ model of a similar size, our estimate with FT-VQE is $10^{4}-10^{5}$ \textit{T}-gates.

\subsection{VQE using the Ross-Selinger (RS) decompositions}

We now evaluate and optimize the convergence of the full VQE minimization using the circuit with the RS decomposed rotation gates. As starting point for the minimization process we use random initial rotation parameters, and compile these into their RS decomposition for a given $d$. We use gradient-based optimizers. To obtain the gradients we use two different methods, and compare the respective convergence behavior. In the first method we evaluate the energy for the parameters shifted by a small finite difference from their previous values (see Sec. \ref{vqe} for details), and we compute the gradient from the resulting energy difference. We choose $\Delta\theta_\mu=0.1$ for this small shift, because the shift needs to be orders of magnitude larger than the error made in the RS decomposition. In general, the precision of gradients that can be obtained using FT($\theta$)-circuits depends on $d$. In the second method we use the parameter-shift rule outlined in Sec. \ref{vqe} to obtain the gradients. Once the gradients are obtained, the optimizer updates the rotation angles as part of the energy minimization process. The updated rotation parameters are again recompiled into FT($\theta$)-circuits. This process is iterated until convergence of the VQE optimization. The optimizer used for both methods is the Broyden–Fletcher–Goldfarb–Shanno (BFGS) algorithm \cite{broyden,fletcher,goldfarb,shanno}. The stopping criterion for the minimization is when the difference 
 energy between the new iteration and the previous one is less than $  10^{-14}$. The same criterion is taken for both the Rz($\theta$)-circuit and the FT($\theta$)-circuits. 

\begin{figure}
    \centering
    \includegraphics[scale = 0.5]{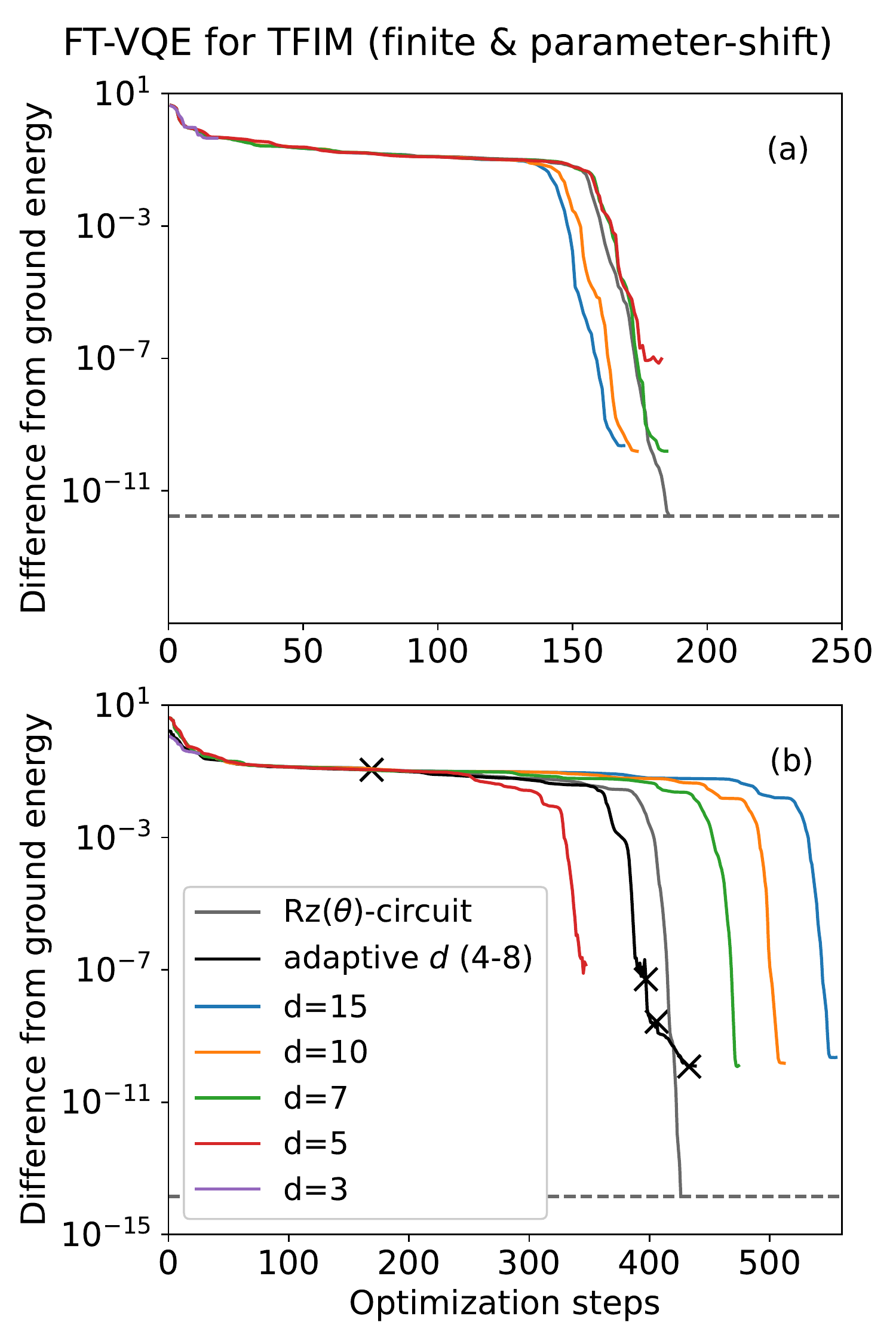}
    \caption{Fault-tolerant VQE with (a) finite-differences with $\Delta\theta_\mu = 0.1$ and (b) parameter-shift rule methods for TFIM. Energy differences were calculated with respect to the numerically exact ground state energy value for the TFIM calculated via matrix diagonalization. The dashed lines in the plots illustrate the optimal expectation value of the Rz($\theta$)-circuit with continuously parameterized rotation gates.}
    \label{fig:tfim_ftvqe}
\end{figure}
We now analyze the convergence behavior of this FT-VQE method by evaluating the energy difference between the energy at each minimization step and the energy obtained by direct numerical diagonalization of the Hamiltonian matrix, which is exact up to numerical precision. This data is obtained from the TensorFlow Quantum dataset available at \cite{tfq}. 
Fig. \ref{fig:tfim_ftvqe} shows the results for the TFIM model, for both the finite-difference based gradient (Fig. \ref{fig:tfim_ftvqe}(a)) and for the gradient obtained using the parameter-shift rule (Fig. \ref{fig:tfim_ftvqe}(b)). We also show the convergence behavior for the Rz($\theta$)-circuit (grey solid line) as reference. Both in Figs.\ \ref{fig:tfim_ftvqe}(a) and (b) we observe that the convergence behavior of the Rz($\theta$)-circuit and FT($\theta$)-circuits is similar. Up to about 150-300 optimization steps the energy decreases very slowly, and then rather abruptly the decrease of energy becomes much larger until it converges. This initial slow convergence is caused by small gradients in large parts of the energy landscape, which are commonly referred to as barren plateaus. 
It can be seen upon close inspection that the light purple $d=3$ plot gets stuck in the barren plateau, where the energy error remains high at a value of the order of $10^{-1}$. The gradient in the barren plateau is characterized by its small magnitude, which leads to correspondingly small updates in the optimization process. In this case, a higher $d$ is needed to update the angle with enough accuracy. For $d\geq5$, sufficiently high accuracy has been reached allowing the convergence of the VQE. For $d\ge5$ the convergence behavior in presence of barren plateaus is similar for both the Rz($\theta$)-circuit and FT($\theta$)-circuits.

We note that the finite-differences method escapes the barren plateau at around 150 steps, while for the parameter-shift rule this happens above 300 steps. We attribute this behavior to large shift of $\Delta\theta_\mu = 0.1$ in the angles for the gradient calculation. The rather large $\Delta\theta_\mu$ can lead to random errors of that order in the gradient. A stochastic component during optimization resulting from such random errors in finite precision gradients can speed up the convergence by bringing the system out of a barren plateau. These random errors can also help the system escape out of a local minimum and hence improve convergence.
The Rz($\theta$)-circuit converges to an error of $10^{-12}$ with the finite-shift rule and to $10^{-14}$ with the parameter-shift rule. On the other hand, the FT($\theta$)-circuits converge to a similar error level of $10^{-10}-10^{-9}$ with both finite-differences and parameter-shift rule for $d\ge7$. FT($\theta$)-circuits use RS decompositions, and as a result involve many more multiplications of matrices as compared to the Rz($\theta$)-circuits. The higher number of multiplications amplifies the numerical precision errors in Qulacs matrix multiplications. This is consistent with the results of Fig. \ref{fig:t_depths}, and with the numerical precision analysis presented in Appendix \ref{mathematica}.

As explained in section \ref{vqe}, in the finite-shift rule one has to recompute the RS decompositions for the shifted angles, whereas in the parameter-shift rule the required angle shifts can be realized with $S$-gates. Therefore, in this setting the parameter-shift rule has less classical overhead for the computation of the gradient. Furthermore, since the required shifts are multiples of the $S$-gate, the gradient calculations do not contribute to the $T$-count. Finally, due to its resilience to noise, as outlined in Ref. \cite{qml_review}, overall we expect that the parameter-shift rule is more efficient for FT-VQE.

\begin{figure}
    \centering
    \includegraphics[scale=0.31]{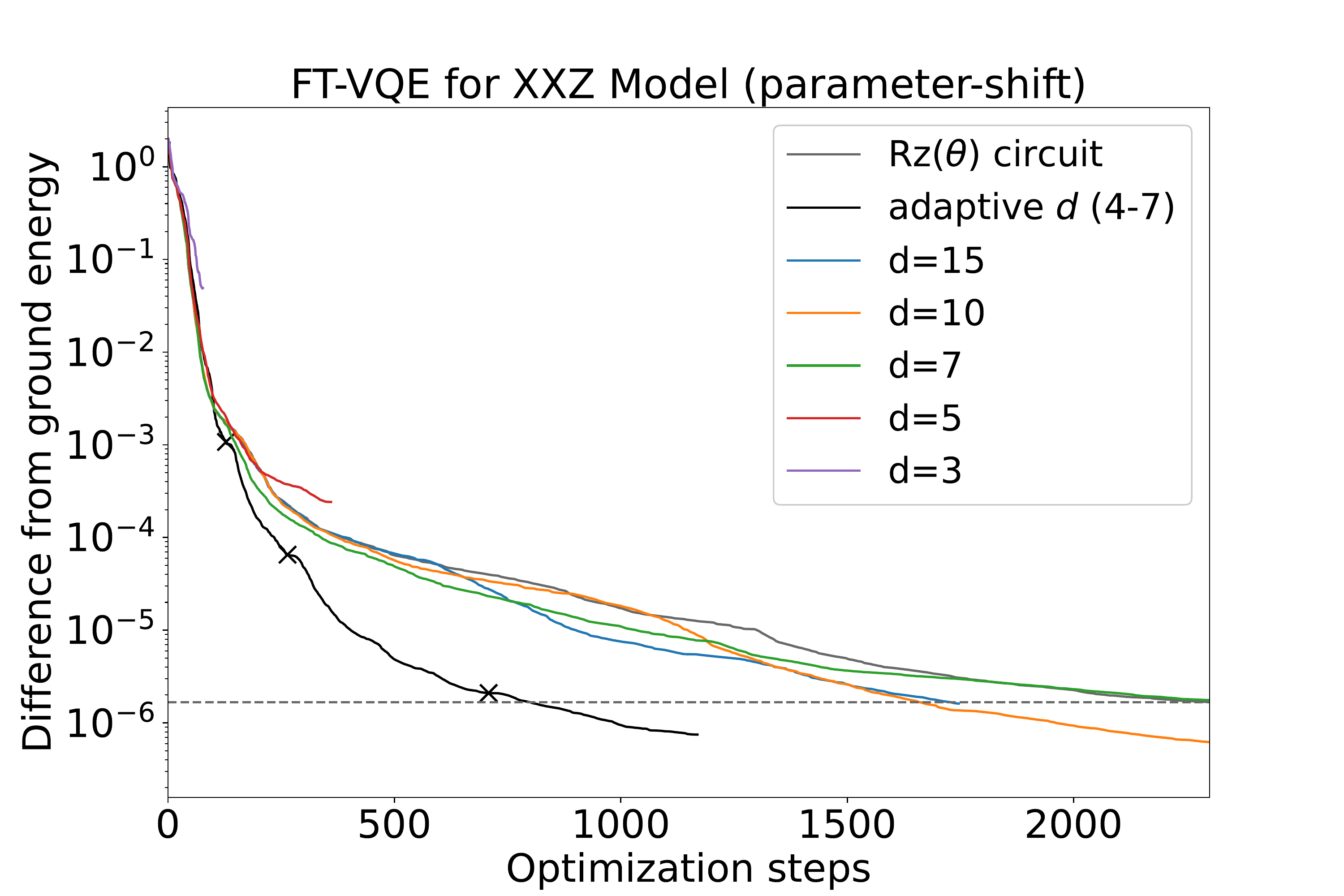}
    \caption{Fault-tolerant VQE with parameter-shift rule method for the XXZ model. Energy differences were calculated with respect to the numerically exact ground state energy computed using matrix diagonalization. The dashed line illustrates the optimal expectation value of the Rz($\theta$)-circuit with continuously parameterized rotation gates.}
    \label{fig:xxz_ft_vqe}
\end{figure}

Fig. \ref{fig:xxz_ft_vqe} shows the convergence behavior for the XXZ model, where we only use the parameter-shift method. One can see that there are no barren plateaus, differently to what was found for the TFIM system. One possible reason for this can be the larger number of variational parameters in the XXZ model ansatz. Such over-parameterization can help avoid barren plateaus, as discussed in Ref. \cite{hva_new}. The system converges rather rapidly to errors below $10^{-2}$ and $10{^{-3}}$, which can be typical accuracy thresholds for practical applications. While for $d=3$ the system only converges down to $10^{-1}$, for $d=5$ one reaches an accuracy of below $10^{-3}$, and hence  $d=5$ can be the appropriate digit precision of the RS decomposition for practical applications.  
If one further increases $d$ and allows for more optimization steps, the energy difference keeps decreasing down to values of $10^{-6}$ and below. Importantly, the FT($\theta$)-circuit convergence at these high $d$ values is very similar to the one for the Rz($\theta$)-circuit. Note that even the Rz($\theta$)-circuit ansatz has an inherent accuracy limit compared to the ground energy calculated from matrix diagonalization, which is the reason for the small but finite remaining energy difference. 

The results show that the small values of $d$ can achieve convergence to moderate errors in the energy, and that with increasing $d$ one can systematically improve the accuracy of the final energy up to a numerical accuracy limit determined by the precision of the software. Furthermore the results also show that when barren plateaus are encountered, random errors in the gradients can help the system escape from the barren plateau. For example, in Fig. \ref{fig:tfim_ftvqe}(b) the $d=5$ FT($\theta$)-circuit escapes the barren plateau at significantly lower optimization step number when compared to larger $d$. We also note that a priori it is not known which $d$ will allow the system to reach a target accuracy. We therefore propose an adaptive algorithm, where a small $d$ is used to start the minimization process, which is then progressively increased during the convergence until a target accuracy is reached. We implement this adaptive approach in our software, where we minimize the energy for a given $d$ until the criterion $\abs{\mathcal{E}(\boldsymbol{\theta}_{t+1})-\mathcal{E}(\boldsymbol{\theta}_{t})}= 1\times 10^{-14}$ is reached. Once this is the case, we proceed by increasing the accuracy $d$ by one.
We show the results for the TFIM model in Fig. \ref{fig:tfim_ftvqe}(b), and for the XXZ model in Fig. \ref{fig:xxz_ft_vqe} (black curves). We set the initial $d$ to $d=3$, and then allow $d$ to increase up to a maximum of $d=8$ within the adaptive algorithm. Each `$\times$' marker represents the point where $d$ is incremented. The results show that the adaptive approach leads to efficient convergence, while allowing to minimize the \textit{T}-depth of the circuit. We therefore expect a generally improved performance of adaptive FT-VQE when compared to FT-VQE at fixed $d$.

\section{Conclusions}\label{conclusions}
Our results demonstrate that variational quantum algorithms, such as VQE, show promise for practical applications on error-corrected quantum computers with limited \textit{T}-depth. Our findings suggest that VQE is viable for running on a fault-tolerant quantum computer. We expect that FT-VQE alone cannot achieve quantum advantage, but that it will be a required component of a number of quantum algorithms that combine different methods. For  example, VQE can be the first step of an algorithm to compute the ground state, where VQE is used to prepare a state with finite overlap with the ground state, which is then used as a starting state for other algorithms such as quantum phase estimation.

We presented the FT-VQE algorithm, which we demonstrated on a quantum emulator for 12 and 16 qubits for two prototypical spin systems. Our study shows that FT-VQE convergence behavior is analogous to standard VQE, especially when we use our proposed adaptive setting of the Ross-Selinger circuit re-compilation accuracy.

Our results highlight the potential of FT-VQE as a powerful tool for practical quantum applications, with particular relevance in quantum chemistry simulations and optimization problems. These findings contribute to the ongoing development of quantum computing technologies and underscore the importance of continued research in this area.

\begin{acknowledgments}
HS is supported by the Engineering and Physical Sciences Research Council [grant number EP/S021582/1]. HS also acknowledges support from the National Physical Laboratory. HS acknowledges the use of the UCL Myriad High Performance Computing Facility (Myriad@UCL), and associated support services, in the completion of this work.
FJ, AA, and IR acknowledge the support of the UK government department for Business, Energy and Industrial Strategy through the UK National Quantum Technologies Programme.
DEB acknowledges the support of the Engineering and Physical Sciences Research Council [grant numbers EP/S005021/1 and EP/T001062/1] and InnovateUK.

\end{acknowledgments}
\newpage
\bibliography{main}

\appendix
\section{Numerical precision of the classical emulation}
\label{mathematica}
To evaluate the numerical precision of the FT-VQE circuits we first evaluate the precision of the RS circuit synthesis for a single rotation gate, and then evaluate how the implementation of a sequence of gates in the used emulators affects the overall numerical precision.

\subsection{Ross-Selinger error analysis}
We carry out error analysis for the Ross-Selinger algorithm in order to assess whether the $z$-rotation realized by the approximate Clifford+\textit{T} unitary $U$ is close to the expected rotation $\theta$, and whether the error is compatible with the operator norm given in Eq.\ \ref{eq:norm}. Furthermore, the non-zero off-axis rotation realized by the unitary $U$ is also calculated. For this purpose, we consider the Euler angle decomposition of a general SU(2) matrix in the ZXZ convention, given by
\begin{align}\label{eq:euler}
\begin{split}
    U(\theta_1,\phi,\theta_2) = e^{-i\frac{\theta_1}{2}Z}e^{-i\frac{\varphi}{2}X}e^{-i\frac{\theta_2}{2}Z}
    \\ = \mqty[e^{-i(\frac{\theta_1+\theta_2}{2})}\cos(\frac{\varphi}{2}) & -ie^{i(\frac{\theta_1-\theta_2}{2})}\sin(\frac{\varphi}{2}) \\ 
    -ie^{-i(\frac{\theta_1-\theta_2}{2})}\sin(\frac{\varphi}{2}) & e^{i(\frac{\theta_1 + \theta_2}{2})}\cos(\frac{\varphi}{2})],
\end{split}
\end{align}
 We calculate the $z$-rotation angles $\theta_1$ \& $\theta_2$ and the off-axis $x$-rotation $\varphi$. 

\begin{figure}[H]
    \centering
    \includegraphics[scale = 0.33]{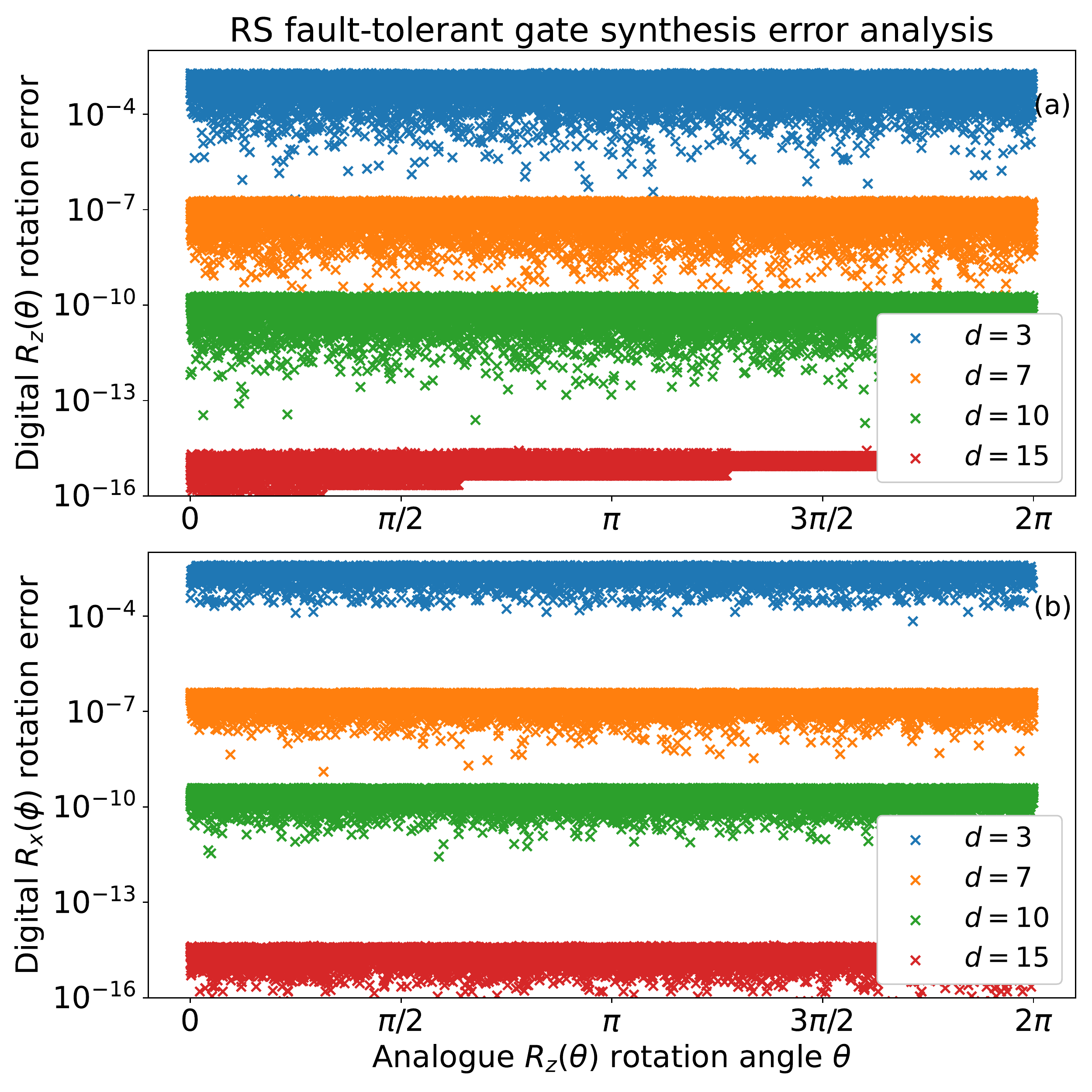}
    \caption{Error analysis of the Ross-Selinger Clifford+\textit{T} circuit synthesis for $R_z(\theta)$, and for $\theta\in[0,2\pi]$: (a) difference between the actual and expected $z$-rotation angle; (b) the off-axis $x$-rotation, which should ideally be $0$. The 
 obtained numerical errors are in line with the expected value set for each digit accuracy $d$.}
    \label{fig:RS_errors}
\end{figure}

In the ideal case, $\theta$ in Eq.\ \ref{eq:norm} should be $\theta\approx\theta_1+\theta_2$, whereas the off-axis rotation should be $\varphi\approx0$. Fig. \ref{fig:RS_errors} shows the results for the computed angles. Note that we take the expression of $U$ for different angles directly from the Ross-Selinger software, so that the only error in consideration is from the RS approximation of the $R_z(\theta)$ rotation. We compute the absolute difference of the Euler angles of $U$ from their expected value. In Fig. \ref{fig:RS_errors} it can be seen that both the $z$-rotations and the undesired $x$-rotations have similar levels of errors, which can be systematically reduced by increasing $d$. 

\subsection{Circuit emulation at finite numerical precision}
Ross-Selinger decompositions lead to a large number of Clifford+\textit{T} gates. When using an emulator, these are implemented as classical matrix multiplications. Due to the large number of matrix multiplications, finite numerical precision errors accumulate and become observable. This behavior can be seen in Figs. \ref{fig:t_depths}(a) and \ref{fig:tfim_ftvqe}(a)(b). To analyze accumulation of precision errors, and to explain why the error levels off when $d$ is increased above some threshold in the FT-circuit emulator runs, we implement an emulator in Mathematica, which allows to perform the computations at arbitrary numerical precision.

\begin{figure}[H]
    \centering
    \includegraphics[scale=0.32]{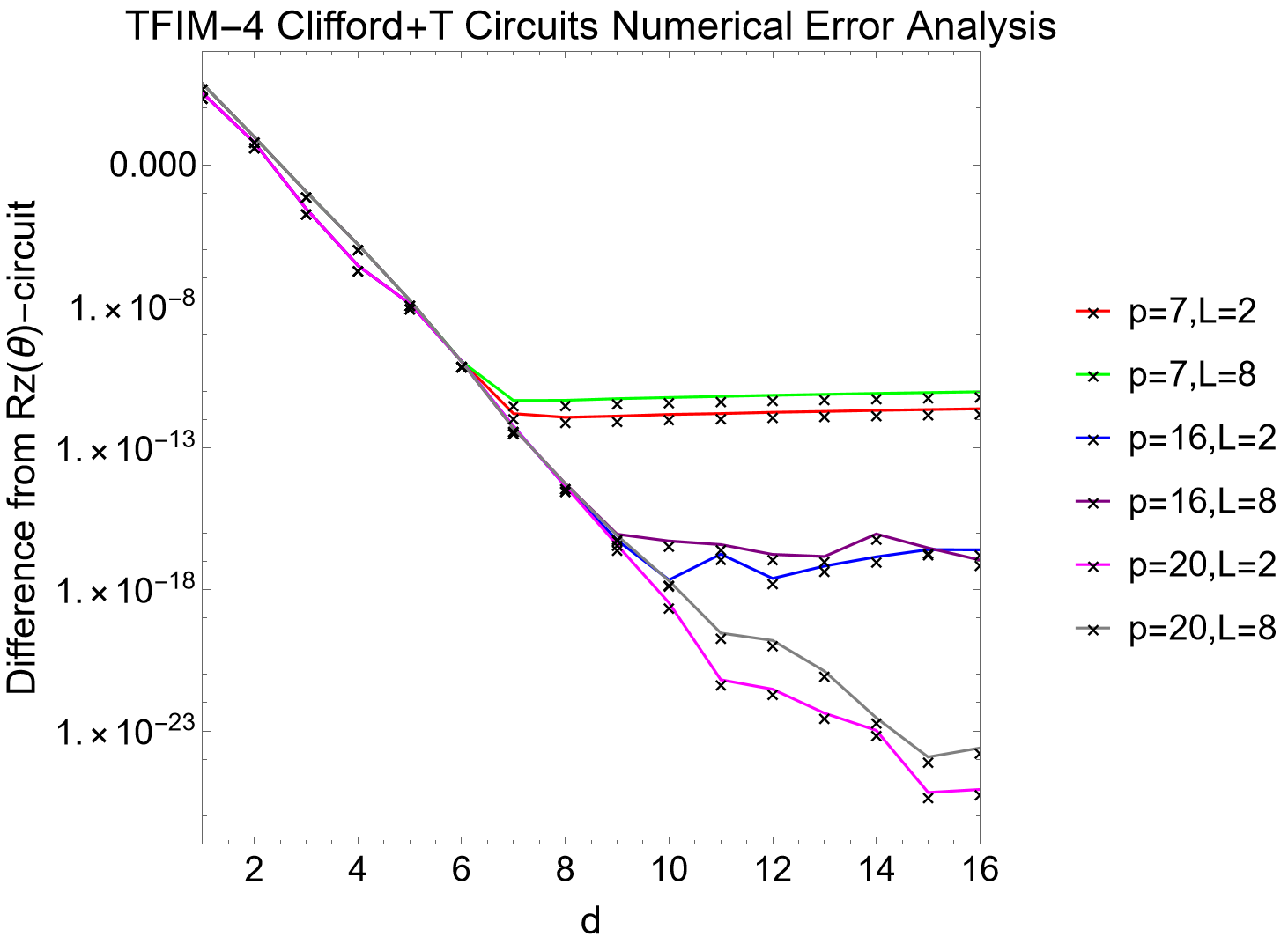}
    \caption{Classical emulation of Ross-Selinger FT-circuit synthesized for increasing accuracy $d$, performed at different levels of numerical precision of the emulator. The model used is TFIM with $N=4$ qubits. $p$ represents the finite numerical precision used in the emulator operations, and $L$ is the layer depth of the TFIM-Q4 ansatz. In the $L=2$ ansatz, there are a total of $16$ parameterized gates with $4$ distinct parameters, whereas in $L=8$, there are $64$ parameterized gates with $16$ distinct parameters.}
    \label{fig:mathematica_error}
\end{figure}

Since the arbitrary precision calculations are computationally expensive for large number of qubits, we use a scaled-down version of the TFIM model. We find that for $N=4$ qubits we can observe the same leveling off of the error with increasing $d$ as for the larger systems. We therefore use this scaled-down system as our test system. In analogy to Fig. \ref{fig:t_depths}, we first calculate the expectation value of the $R_z(\theta)$-circuits with optimized parameters. Then, we perform the RS circuit recompilation for the $R_z(\theta)$-circuits at various $d$, and compute the expectation values. The results are presented in Fig. \ref{fig:mathematica_error}. The axes of Fig. \ref{fig:mathematica_error} and Fig. \ref{fig:t_depths}(a) are the same.

We take two different layer depths of the TFIM 4-qubit ansatz in order to investigate the effect of increasing number of matrix multiplications at various numerical precision values, which we indicate by $p$. In the $L=2$ ansatz, there are a total of $16$ parameterized gates, with $4$ distinct parameters, whereas for $L=8$ there are $64$ parameterized gates, with $16$ distinct parameters. At the limited $p=7$ precision, a similar pattern as that of Fig. \ref{fig:t_depths}(a) is observed, where the error plateaus despite increasing $d$. Importantly, when we increase the precision of our Mathematica emulator to $p=16$ and $p=20$, the leveling off threshold of the error systematically decreases. For $p=20$ no leveling off is observed. This confirms that the leveling off of the error in the emulator runs for the FT-circuits is due to the inherent numerical precision of the emulator.

\end{document}